\documentclass[twocolumn,superscriptaddress,floats,prd,nofootinbib,showpacs]{revtex4-2}

\usepackage{amssymb,amsmath,amssymb,amsfonts,amsthm,stmaryrd,mathrsfs,physics,bbm,mathtools,lipsum}
\usepackage{graphicx}
\usepackage[T1]{fontenc}
\usepackage{enumerate} 
\usepackage{color} 
\usepackage[caption=false]{subfig}
\usepackage{bm} 
\usepackage{hyperref} 
\usepackage[normalem]{ulem}

\newcommand{\tc}{\tau}
\newcommand{\rc}{\tilde r}
\newcommand{\thc}{ \theta}
\newcommand{\phc}{\phi}

\begin{document}

\title{Quantum Oppenheimer-Snyder and Swiss Cheese  models}

\author{Jerzy Lewandowski}
\email{jerzy.lewandowski@fuw.edu.pl}
\affiliation{Faculty of Physics, University of Warsaw, Pasteura 5, 02-093 Warsaw, Poland}

\author{Yongge  Ma}
\email{mayg@bnu.edu.cn}
\affiliation{Department of Physics, Beijing Normal University, Beijing 100875, China}

\author{Jinsong Yang}
\email{jsyang@gzu.edu.cn}
\affiliation{School of Physics, Guizhou University, Guiyang 550025, China}
\author{Cong Zhang}
\email{czhang@fuw.edu.pl}
\affiliation{Faculty of Physics, University of Warsaw, Pasteura 5, 02-093 Warsaw, Poland}
\affiliation{Department Physik, Institut f\"ur Quantengravitation, Theoretische Physik III, Friedrich-Alexander-Universit\"at Erlangen-Nürnberg, Staudtstra{\ss}e 7/B2, 91058 Erlangen, Germany}

\begin{abstract}
By considering the quantum Oppenheimer-Snyder model in loop quantum cosmology, a new quantum black hole model whose metric tensor is a suitably deformed Schwarzschild one is derived. The quantum effects imply a lower bound on the mass of the black hole produced by the collapsing dust ball. For the case of larger masses where the event horizon does form, the maximal extension of the spacetime and its properties are investigated. 
 By discussing the opposite scenario to  the quantum Oppenheimer-Snyder, a quantum Swiss Cheese model  is obtained with a bubble surrounded by the quantum universe. This model is analogous to black hole cosmology or  fecund universes where the big bang is related to a white hole. Thus our models open a new window  to cosmological phenomenology. 
\end{abstract}

\maketitle

According to Subrahmanyan Chandrasekhar "The black holes of nature are the most perfect macroscopic objects there are in the Universe" \cite{chandrasekhar1984mathematical}. Given development of quantum models describing spacetime filled with dust, the following two questions are addressed in this letter: What is a black hole (BH) spacetime containing a collapsing matter ball  like?  What is a BH spacetime  surrounded by  a universe  like?

In classical general relativity (GR),  our understanding on these two  questions is shaped by the the Oppenheimer-Snyder model \cite{oppenheimer1939on}, which depicts the collapse of the pressureless homogenous dust coupled to the Friedmann–Lema\^itre–Robertson–Walker metric. However, this metric appears to be problematic due to the  Big-Bang singularity.  A proposal to resolve this singularity is to replace the Big Bang by a Big Bounce, which was largely considered by cosmologists for aesthetic reasons \cite{kragh1999cosmology}.  
Thus, it is desirable  to answer the above two questions by considering collapsing and bouncing matters.

 Quantum gravity has always been expected to go beyond the singularities of the classical 
GR. Indeed, the existence of a Big Bounce resolving the Big Bang singularity  has found a diverse support in the Loop Quantum Cosmology (LQC) models (see, e.g., \cite{ ashtekar2006quantum,yang2009alternative,assanioussi2018emergent}).
A concrete bouncing model is the Ashtekar-Pawlowski-Singh (APS) model, where the bounce is a rigorous result of  the fundamental discreteness \cite{ashtekar2006quantum}. In this model, the semiclassical metric tensor has the form
\begin{equation}\label{eq:dustmetric}
\dd s^2_{\rm APS}=-\dd \tc{}^2+a(\tc)^2(\dd\rc{}^2+\rc{}^2\dd\Omega^2),
\end{equation}
where  $(\tc,\rc,\thc,\phc)$ denotes a coordinate system,  $\dd \Omega^2=\dd\thc^2+\sin^2\thc\dd\phc^2$. The function $a(\tau)$ satisfies a deformed Friedmann equation
\begin{equation}\label{eq:LQCfried}
\begin{aligned} 
H^2:=&\left(\frac{\dot{a}}{a}\right)^2=\frac{8\pi G}{3}\rho\left(1-\frac{\rho}{\rho_c}\right),\ \rho= \frac{M}{\frac{4}{3}\pi{\tilde r}_0^3 a^3},
\end{aligned}
\end{equation}
where the deformation parameter is  the critical density $\rho_c=\sqrt{3}/(32\pi^2 \gamma^3 G^2\hbar)$ with the Barbero-Immirzi parameter $\gamma$, $M$ is the mass of the ball of the dust with the radius $ a(\tau) {\tilde r}_0$ in the APS spacetime. Note that  the second equation in \eqref{eq:LQCfried} with a constant $M$  is the consequence of the conversion law $\nabla_\mu T^{\mu\nu}=0$ with $T_{\mu\nu}=\rho(\tau) (\nabla \tc)_\mu ( \nabla \tc)_\nu.$  Eq. \eqref{eq:LQCfried}   reverts to the usual Friedmann equation in the classical regime when $\rho\ll \rho_c$. However in  the quantum regime where $\rho$ is  comparable with  $\rho_c$, the equation prevents  $\rho(\tau)$ from reaching infinity. 
This property ensures that the metric tensor  $\dd s^2_{\rm APS}$ is nowhere and never singular. The function $a(\tau)$ can be extended to the whole interval $(-\infty,\infty)$.

The particles of the dust in the APS spacetime (\ref{eq:dustmetric}) are the geodesics satisfying  $\tilde r,  \theta, \phi = \rm const$.  Therefore, an APS dust ball can be characterized as a region $0\le {\tilde r} \le {\tilde r_0}$ of the APS spacetime. Then, our quantum (or rather semiclassical)  Oppenheimer-Snyder (qOS) model assumes the (pseudo) static \footnote{By pseudo static, we take into account the case that the Killing vector $\partial_t$  inside the BH is space-like. } spherically symmetric metric 
\begin{equation}\label{eq:exteriormetric}
\dd s^2_{\rm MS}=-(1-F(r))\dd t^2+(1-G(r))^{-1}\dd r^2+r^2\dd\Omega^2,
\end{equation}
with some functions $F(r)$ and $G(r)$,  where $(t,r,\theta,\phi)$ are coordinates.  The coordinates $\theta$ and $\phi$ are joint for the ball region and the exterior (meaning they are extensions of each other), whereas the coordinates $\tau, \tilde{r}$ are used in the ball region only, while the coordinates $t,r$ are used only in the exterior region. 
Eq. \eqref{eq:exteriormetric} is a minimal assumption if we are to obtain an exact BH metric by the junction  condition, without employing equations of motion. As we are going to show, there are close ties between the models of quantum BH and quantum cosmology, providing a possibility to detect the quantum effects in early universe from BHs.
Actually, the resulting metric is a suitably deformed Schwarzschild one, where the deforming term leads to a BH mass gap. Moreover, the deformed Schwarzschild metric induces a non-vanishing effective energy-momentum tensor. This may relate the quantum effects of BHs with the dark matter.

The  methodology of LQC  could also be applied to  BH models \cite{modesto2004disappearance,ashtekar2005black,gambini2008black,agullo2008black,haggard2015quantum,christodoulou2016realistic,corichi2016loop,ashtekar2018quantum,gambini2020spherically,zhang2020loop,gan2020towards,song2021entropy,assanioussi2021loop,han2022improved,zhang2021reduced,giesel2021nonsingular,husain2022quantum}. However, the answers do not form a unique picture. 
Particularly, according to certain conjecture, the bouncing interior of the  BH destroys the Killing horizon in the future, and the process takes the form of BH evaporation \cite{ashtekar2005black}. According to another proposal, the reflecting interior turns a BH into a white hole (WH), and the transitional region of spacetime is strictly quantum, giving the process the character of quantum tunneling \cite{haggard2015quantum,christodoulou2016realistic}. In both of these cases, the global structure of the null infinity is similar, there is one scri. Subsequent models describe spacetime containing a quantum BH differently (see, e.g.,\cite{ashtekar2018quantum}).  According to them, the bouncing interior does not affect the global structure of the exterior.   Hence, the spacetime  looks similar to the Kruskal diagram of Schwarzschild spacetime, with the only difference that the singularity becomes an edge of spacetime on which the metric is still regular. Now the extension consists in gluing the diagrams, even of an unbounded number, with the edges.  

In the case of the second question,  we consider the opposite scenario: a spherically symmetric, static empty region of spacetime (a bubble) surrounded by the quantum universe according to the APS model. Precisely, this scenario consists in removing the ball  $0\leq \rc\leq \rc_0$ from the APS spacetime, that is considering the APS metric tensor $\dd   s^2_{\rm APS}$ for 
${\tilde r}  \ge {\tilde r_0}$. The hole left by the ball is filled with a piece of the spacetime (\ref{eq:exteriormetric}). 
Hence this is a quantum Swiss Cheese (qSC) model whose physical meaning  is quite different from the qOS model. Before the quantum universe bounces,  the spherically symmetric bubble is being squised. The question is whether its radius shrinks below the radius of the horizon, and if so, what happens next.  Indeed, according to the result shown below, unless the size of the region is of the order of the Planck length, the  horizon does form. The radius depends on  the amount of mass of the dust if it were filling  the bubble. Briefly speaking, the bouncing universe turns the bubbles into BHs.

In both cases the key role is played by the   the dust space  surface (either outer or inner) $
{\tilde r} = {\tilde r_0}$
in the APS spacetime, that in the spacetime (\ref{eq:exteriormetric}) will be described in a partially parametric form $
(t(\tau), r(\tau), \theta,\phi) $
where $ -\infty < \tau < \infty$ is the proper time, and the ranges of coordinates read $0\le \theta\le \pi, \ 0\le \phi< 2\pi$.  We glue the spacetimes by the identification $(\tau,{\tilde r}_0,\theta,\phi) \sim (t(\tau),r(\tau),\theta,\phi)$
such that the induced metric and the extrinsic curvature are equal on the gluing surfaces that become a single surface of the dusty part of the spacetime. That will allow  us to unambiguously determine the functions $F$ and $G$  as well as a location of the dust surface in the dust-free spacetime --- asymptotically for $r\rightarrow \infty $ it is tangent to $\partial_t$.  Then, the metric  \eqref{eq:exteriormetric}  can be obtained as
\begin{equation}\label{eq:exteriormetric1}
\begin{aligned}
\dd s^2_{\rm MS}=&-\left(1-\frac{2GM}{r}+\frac{\alpha G^2 M^2}{r^4}\right)\dd t^2\\
&+\left(1-\frac{2GM}{r}+\frac{\alpha G^2 M^2}{r^4}\right)^{-1}\dd r^2+r^2\dd\Omega^2,
\end{aligned}
\end{equation}
where we introduced  the parameter $\alpha=16\sqrt{3}\pi\gamma^3\ell_p^2 $ with $\ell_p=\sqrt{G\hbar}$ denoting the Planck length. Indeed, the calculation to determine the functions $F$ and $G$  is quite straightforward (see  Appendix \ref{App} for details). 
 It is worth noting  that  the form \eqref{eq:exteriormetric1}  of the metric is determined for
\begin{equation}\label{eq:rb}
r\ge r_{\rm b}=\left(\frac{\alpha GM}{2}\right)^{\frac{1}{3}},
\end{equation}
which results from the fact that the dust surface radius $a(\tau)\rc_0$  runs over $[r_{\rm b},\infty)$. Hence the functions $F(r)$ and $G(r)$  may be defined arbitrarily for $r<r_{\rm b}$ . The parameter  $M$  coincides with the ADM mass of the metric tensor \eqref{eq:exteriormetric1}. As a quantum deformation of the Schwarzschild metric, the spacetime  metric tensor \eqref{eq:exteriormetric1} coincides with that derived in \cite{kelly2020effective,parvizi2022rainbow,loops22}.  

 The global structure of the spacetime determined by \eqref{eq:exteriormetric1}  depends on the number of roots of $1-F(r)$. It is convenient to introduce the parameter   $0<\beta<1$ by 
\begin{equation}
G^2M^2= \frac{4 \beta ^4}{\left(1-\beta ^2\right)^3} \alpha.
\end{equation}
For $0<\beta<1/2$,  that is when 
\begin{equation}\label{Min}
M < M_{\rm min}:=\frac{4}{3\sqrt{3}G}\sqrt{\alpha},
\end{equation}
$1-F(r)$ has no real root, implying that the metric \eqref{eq:exteriormetric1} does not admit any horizon.  The global causal structure of the maximally extended spacetime  is the same as that of the Minkowski spacetime.  Hence the value 
\begin{equation}\label{eq:Min1}
M_{\rm min} = \frac{16\gamma\sqrt{\pi\gamma}}{3\sqrt[4]{3}}\frac{\ell_p}{G}
\end{equation}
is a lower bound for BHs produced by our  models (see  \cite{zhang2022loop,giesel2021nonsingular,husain2022quantum} for compatible results).   The minimal mass is of the order of the Planck mass. Its actual value depends on the value of the Barbero-Immirzi parameter $\gamma$ of LQG, that is argued to be of order of $0.2$ \cite{meissner2004black,domagala2004black}. 

%
%
%

\begin{figure}[h]
\centering
\includegraphics[width=0.4\textwidth]{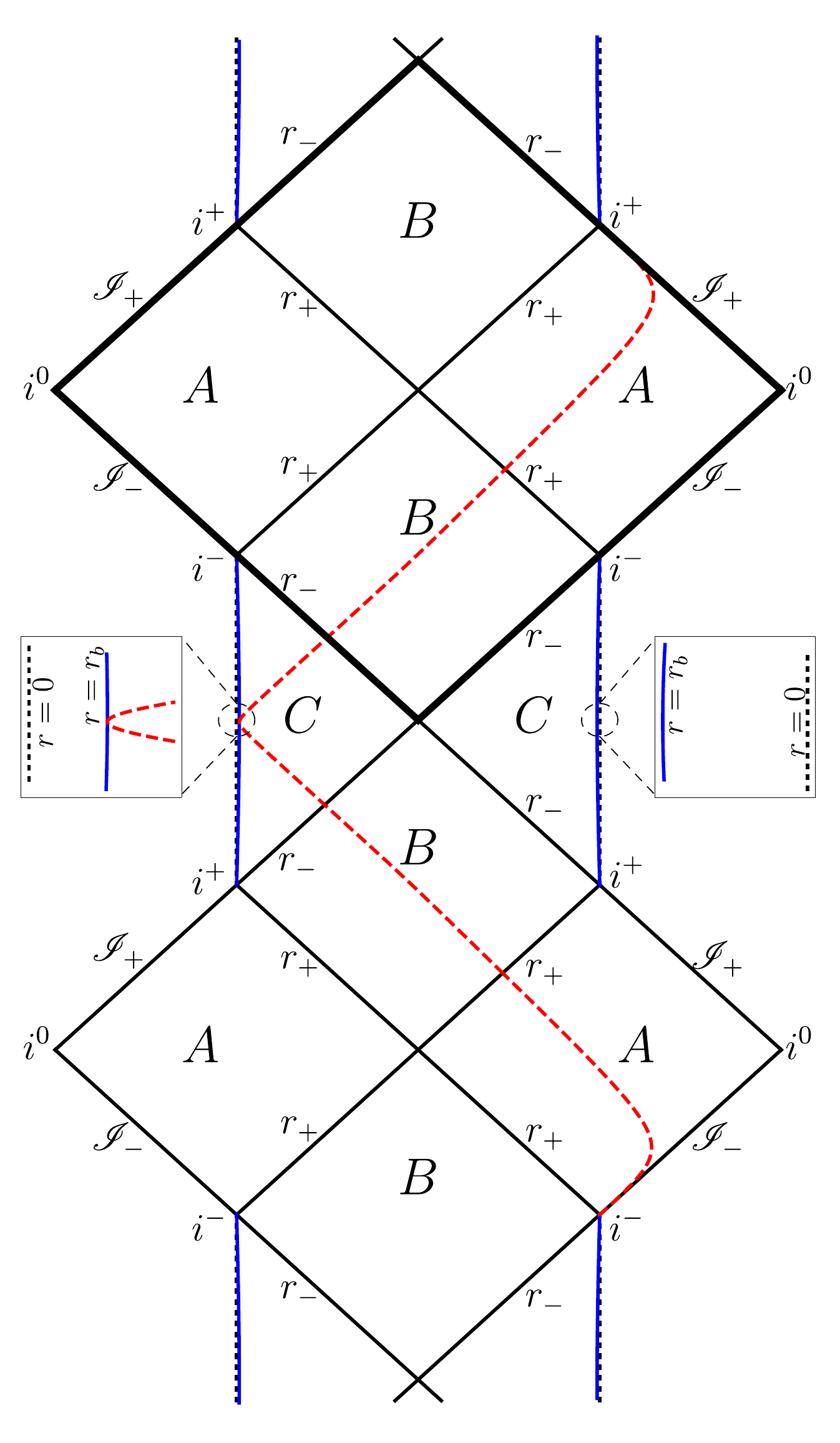}
\caption{Penrose diagram of the maximal extension for $1/2<\beta<1$. The geodesic of the dust surface is plotted in the red dashed line. 
The blue line plots $r=r_b$. Indeed, $r_b$ is the root of $F(r)$. The doted lines plot $r=0$ if we  analytically extend $\dd s_{\rm MS}^2$.  
A modified Kruskal region is encircled by the thick lines. 
}\label{fig:geopen2}
\end{figure}

 Consider the case  of 
$M>M_{\rm min}$, i.e.,  $1/2<\beta<1$.   The function $1-F(r)$ has exactly two roots 
$$r_\pm=\frac{\beta\left(1\pm \sqrt{2\beta-1}\right)}{\sqrt{(1+\beta)(1-\beta)^3}}\sqrt{\alpha},$$
that makes the coordinate $t$ singular.  We extend  the metric tensor  $\dd s^2_{\rm MS}$ by following the steps similar to those for the Reissner--Nordstr\"om (RN) metric. The resulting Penrose diagram  Fig. \ref{fig:geopen2} also has the structure similar to that of the RN   spacetime. The $A$ regions (such that  $r>r_+$) are static and asymptotically flat (even asymptotically simple with both future and past complete scris). The $C$ regions  (such that $0\le  r< r_-$) are static, however,  for $r<r_{\rm b}$, the metric tensor is not determined by the junction conditions. Both the $A$ regions and $C$ regions contain complete orbits of the time translation, where the time variable ranges from $-\infty$ to $\infty$.  Finally, the $B$ regions (such that $r_-<r<r_+$) are nowhere static, they are  trapped or anti-trapped. The surfaces $r=r_\pm$ set bifurcated Killing horizons, whereby branches of the $r=r_+$ horizons are BH / WH event horizons, while branches of the $r=r_-$ horizons are Cauchy future/ past horizons.   

Now we can maximally extend the surface of the dust region that is generated by the geodesics of the dust surface. This will be either the surface of the ball of dust or the inner surface of the dusty universe surrounding a bubble of static (in some regions) spacetime. The surface  emanates from  the most past corner $i^-$ of an $A$ region, crosses the BH horizon, passes  across the $B$ region and crosses the opposite inner horizon (the both branches of the crossed horizons can be covered by a single advanced Eddington-Finkelstein coordinate),  reaches the minimal value $r_{\rm b}$ of the $r$ coordinate, bounces, and continues symmetrically all the way to most future corner $i^+$ of another $A$ region.   

In the degenerate case 
$M=M_{\rm min}$,     the   $B$ regions shrink and the spacetime consists of the static $C$ and $A$ regions, and the  Killing horizons corresponding to $r_-=r_+$ are not bifurcated  but become BH / WH event horizons.  As the case of $M>M_{\rm min}$, the Penrose diagram has the structure similar to that of the extremal RN spacetime \cite{hawking1973large}.

\begin{figure}[h]
\centering
\includegraphics[width=0.22\textwidth]{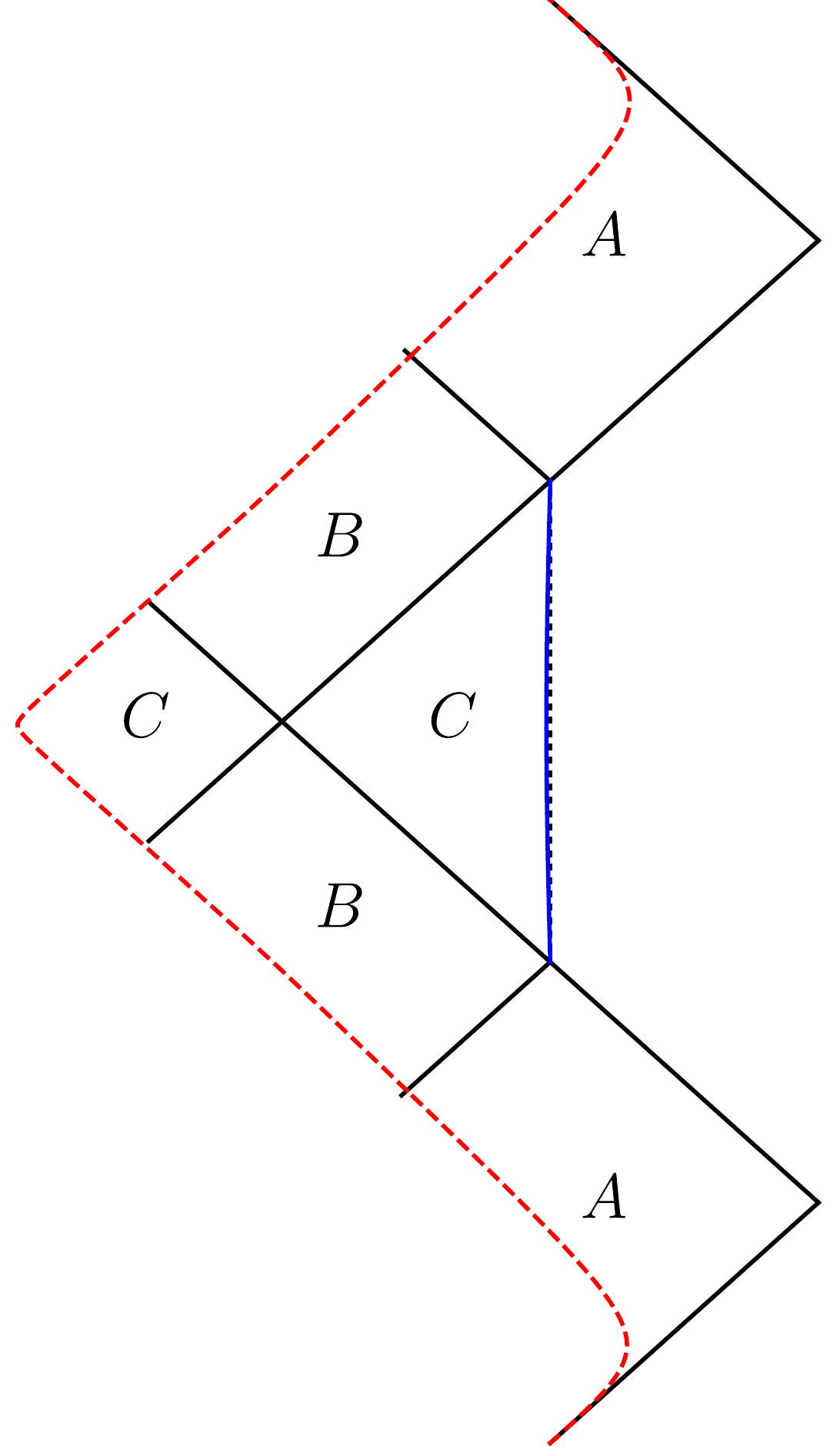}(a)
\includegraphics[width=0.2\textwidth]{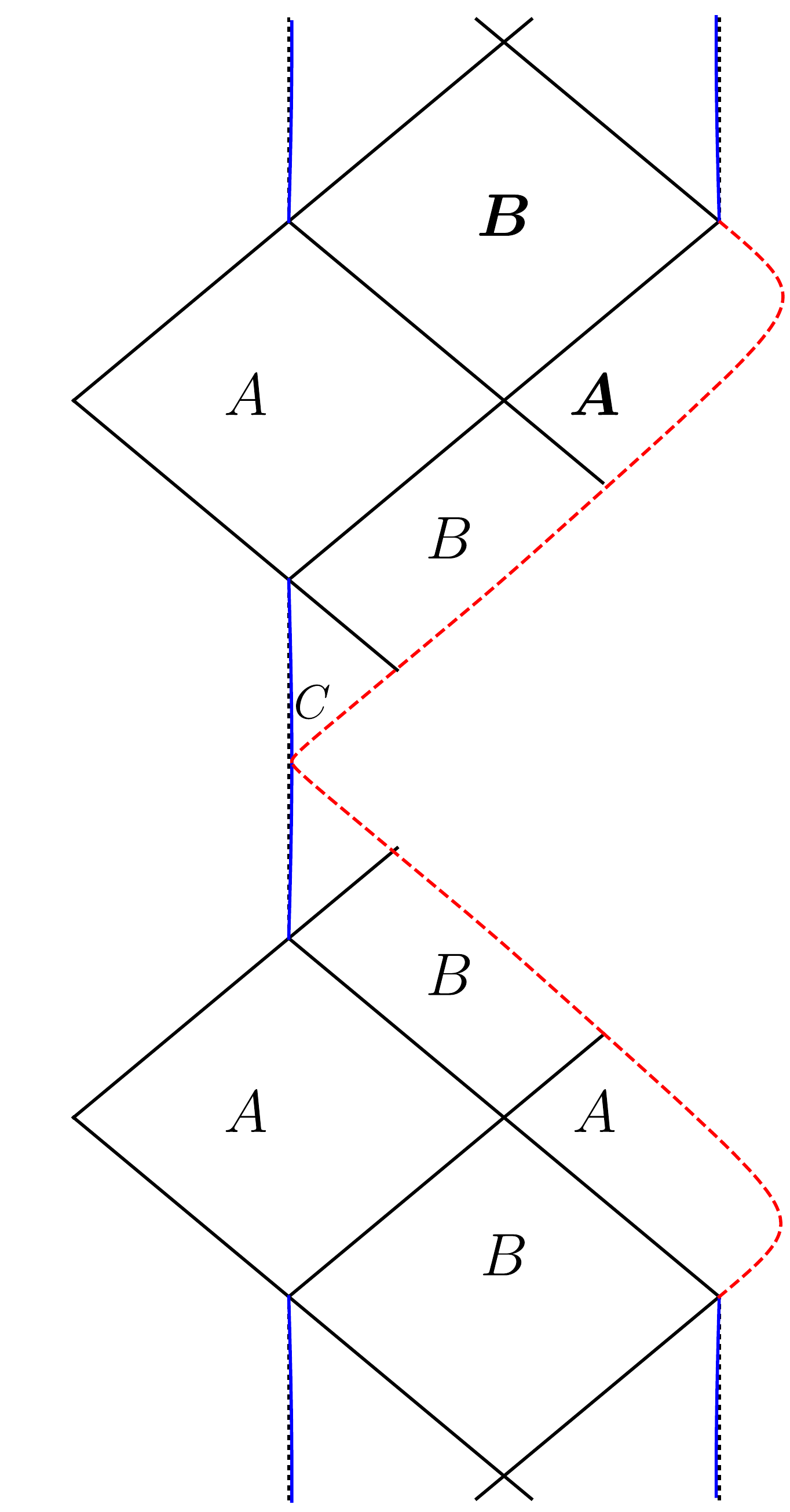}(b)
\caption{(a) The piece outside the collapsing dust of the Penrose diagram containing the  collapsing dust, for $1/2<\beta<1$. (b) The piece inside the LQC dust of the Swiss cheese   diagram. The $A$-region labeled by  the bold $A$-letter is the asymptotically  flat region of the current universe in the qSC model, and the one labeled by the bold $B$-letter is the trapped region accessible for observers in the current universe.  
}\label{fig:geopen2p}
\end{figure}

We now turn to our qOS model. In the case $M<M_{\rm min}$, the ball of the APS dust collapses from infinite radius to the radius $r_{\rm b}$ (\ref{eq:rb}), bounces, and expands, and the world sheet of the ball surface is symmetric with respect to the bounce.  The exterior is just a single static, asymptotically flat (and asymptotically simple) region, such that every point of the exterior can be connected with every of the  scris with causal curves. No black or white holes emerge.      

 Suppose $M > M_{\rm min}$.  Replacing the part of the diagram \ref{fig:geopen2}  to the left of the surface, i.e. surface's  interior,  with the spacetime \eqref{eq:dustmetric} of dust ball with $\rc<\rc_0$, we get the Penrose diagram containing the collapsing dust. For clarity, the region of the spacetime outside the collapsing dust  is plotted in Fig. \ref{fig:geopen2p}(a) \cite{loops22}.  We call it exterior. 
 It is contained in  the  $A$  regions, $B$ regions and $C$ regions. Consider the part of the exterior that is  contained in the past $A$ region. It is static and asymptotically simple, however it is bounded not only by the surface of the ball of dust,  but also  by event horizon beyond which the ball disappears.  
 No observer who stays in this part of the exterior will know about the bounce of the ball spacetime and  the expanding phase. They will see collapsing ball that sinks into the horizon. The exterior part contained in the future $A$ region is just the time inverse of the past part.   An observer staying at the site sees first the WH and then a ball of dust pouring out of it. They may receive information about the formation of a BH in the past, but they will never experience it themselves. An outside ball observer in the past $A$ region also has the option of crossing the event horizon. After crossing the non-static region $B$, they will be in one  of the two static regions $C$. From one of them, they can follow the bouncing sphere to the future region $A$. However, there is another $C$ region outside the sphere that is static and exceeds the radius limit $r_{\rm b}$ away from the sphere. There, the observer can enter static areas with values $r<r_{ b}$ while still remaining outside the sphere.  This part of the exterior  is spatially bounded but temporally unbounded, hence the observer may stay there for ever on one of the Killing orbits and never hit the ball.  The WH horizons are at the same time the Cauchy horizons, as in the RN spacetimes. While they are mathematically convincing, physically they are likely to be unstable with respect to perturbations. Indeed, it is easy to find an example of such a contour that a flux of the energy of a radiation through a finite wall transverse to the Cauchy horizon would have to balance the flux  through another wall that terminates in the future scri of the previous region $A$ and is therefore infinite.       

 In the case of $M=M_{\rm min}$  the asymptotically flat  $A$-type static regions of the exterior have the same properties as described above in the  case of a greater $M$. The exterior regions beyond the horizon though are different. Namely, the non-static regions of the $B$  do not occure at all and the $C$-type region has only the part containing the bouncing ball but does not extend to regions of $r<r_{\rm b}$.  Hence, any external observer that entered the BH horizon will get to the future $A$ region in some point. 
 
In conclusion, the exterior observes living in the asymptotically simple past $A$-type regions  may not see any quantum effects in the causal structure. All they will see is a collapsing ball  that disappears beyond a horizon. There is however a quantum effect on the induced metric tensor. If we define the energy-momentum tensor of the metric tensor \eqref{eq:exteriormetric1} by 
$T^{\rm q}_{\mu\nu} := G_{\mu\nu}/(8\pi G)$, 
where $G_{\mu\nu}$ is the Einstein tensor, then we find  that a Killing observer perceives energy density
\begin{equation}\label{eq:rhoq}
\rho^{\rm q} =\frac{3 \alpha  G M^2}{8\pi r^6}.
\end{equation} 
 Clearly, when the  quantum deformation of the APS spacetime vanishes, then  $T^{\rm q}_{\mu\nu} $ vanishes as well, hence it is a purely quantum effect.   

There is one more remark. Even if we just analytically extend the spacetime \eqref{eq:exteriormetric1} to the region $r<r_{\rm b}$ such that 
 the diagram  contains a singularity, i.e., the one at $r=0$ in  Fig. \ref{fig:geopen2p}(a).  This singularity cannot be hit by timelike geodesics, because all timelike geodesics turn out to enter the left $C$ region in Fig. \ref{fig:geopen2p}(a), just like the dust surface geodesic. This means  that the  spacetime is timelike geodesically complete.  In addition, even though there are timelike non-geodesics reaching the singularity, the singularity is still physically inaccessible for observers. This statement results from the fact that those timelike curves reaching the singularity carry infinite integrated acceleration, in  contrast to the finite integrated acceleration along the world line of a physically reasonable observer which carries finite payload \cite{chakrabarti1983timelike}. 

Let us turn to the qSC model.  In this model, we are concerned with the cases where the  horizon does form as the bubble is being squeezed. Thus, we suppose $M\geq M_{\rm min}$. 

For $M>M_{\rm min}$, replacing the part of the diagram \ref{fig:geopen2} to the right of the surface with the spacetime of the dust universe with $\rc>\rc_0$, we get the Penrose diagram of the qSC model.  For clarity, the  region of the spacetime inside the collapsing dust  universe is plotted in Fig. \ref{fig:geopen2p}(b).


 The qSC diagram  \ref{fig:geopen2p}(b) contains  infinitely many $A$, $B$, $C$ regions.  Consider observers staying in the left past $A$ region. In order to not hit the dust universe, the observers have to travel towards the center of the bubble. Then, crossing the horizon at $r_+$, they will enter the trapped region $B$, and pass the horizon $r=r_-$ to get into  the wormhole region $C$. There, they can fall into the region with $r<r_b$ to avoid the dust universe and move into the expansion epoch. In the  expansion epoch, the observers can stay in wormhole region $C$ forever, or follow the expanding dust universe to arrive at the right future $A$ region,  i.e. the  piece of the diagram \ref{fig:geopen2p}(b) labelled by the bold $A$-letter. This region is static and asymptotically simple, and is called the current universe.  Observers living here  can see a WH from the past.  This WH is the same as that in the Kruskal spacetime up  to  some quantum correction. According to this discussion, the qSC relates the Big Bang with a WH, which  could open a new door for the cosmological phenomenology, like a new explanation on the fast radio bursts  and some high-energy cosmic rays \cite{barrau2014fast}. Indeed, there have been a few discussions relating the Big Bang to a WH \cite{pathria1972universe,retter2012revival,aguilar2014primordial}. Our qSC model depicts a picture analogous to the BH cosmology \cite{pathria1972universe} or  fecund universes \cite{smolin1998life}.

Observers in the current universe can also observe a BH horizon that is the same as the Schwarzschild one up to some quantum correction.  This correction could cause some remarkable effects.  At first, a Killing observer could perceives energy density  $\rho^{\rm q}$ given by \eqref{eq:rhoq}. With $\gamma=0.2375$, $\rho^{\rm q}$ around the horizon of a solar mass BH takes value $\sim 10^{17} \rm{kg}/{\rm m}^3$. This could provide a new piece of  the dark matter.  Moreover, once the observes cross the horizon,  their fate will be quite  different from the classical one. Unlike in the Schwarzschild spacetime where they have to hit the spacelike singularity, they will pass the trapped region, experience an anti-trapped $B$ region after a wormhole $C$ region, and move into another universe. That universe could not yet be their final destination, as they can enter another trapped $B$ region through that universe and continue  their journey.

To summarize,
 the (pseudo) static spherically symmetric spacetime (\ref{eq:exteriormetric1})  that contains a collapsing ball of quantum APS dust  (the flat model) is determined as a suitably modified Schwarzschild spacetime.   The only assumption is that  the metric tensor is at least first-order differentiable at the junction surface. 
 There is a lower bound $M_{\rm min} = \frac{16\gamma\sqrt{\pi\gamma}}{3\sqrt[4]{3}}\frac{\ell_p}{G}$ for the mass in order to create a BH.  For the larger $M$, the observers sitting in a past asymptotically flat region will see the BH formed by the collapse but will never see the ball bounce behind the horizon. However, they will feel a non-zero energy-tensor induced  by the quantum properties of the dust ball. That apparent matter curves the spacetime, and hence it  has properties of the  dark matter. The future  asymptotically flat region is the time reflection.  The WH horizon therein is a Cauchy horizon that is (most likely) unstable with respect to perturbations of spacetime by the analogy with the RN spacetime. The spacetime metric \eqref{eq:exteriormetric1} is not determined for $r$ less than the radius of the bounce of the ball. But even if we consider  just the analytic continuation, the singularity inside is timelike and reaching it would take infinite energy (similarly to  the RN case).   

If a bubble of the (pseudo) static spherically symmetric spacetime  is surrounded by a universe of the quantum APS dust, then again the metric tensor is determined as the  modified Schwarzschild metric. The same lower bound for the emergence of the horizon applies. Probably the most important quantum effect is the emergence of the apparent matter that really curves the spacetime. The future static part of the bubble contains both the BH and the WH.  After jumping inside the BH horizon an observer has options similar to those in the RN spacetime. The new pictures provided by the two models
open a new window to test the effects of quantum gravity through the cosmological phenomenology.
 
It should be noted that our method to obtain the modified Schwarzschild metric is also valid for more general effective dynamics of the collapsing dust ball.
Given a deformed Friedemann equation $H^2=8\pi G\rho X(\rho)/3$ with a general function $X(\rho)$, following the derivations in Appendix \ref{App}, we get the functions $F(r)$ and $G(r)$ in \eqref{eq:exteriormetric} as  $F(r)=G(r)=2GM r^{-1}X\left(3 M/(4\pi r^3)\right)$. Note also that while the current work concerns the flat LQC model since it is simple and well-understood, the qOS and the qSC models with dust ball governed by open/close LQC dynamics can be investigated similarly. Taking the LQC models with $k=\pm 1$ in \cite{vandersloot2007loop,corichi2011loop,langlois2017effective} for instance and applying the junction condition, one can obtain $F(r)=G(r)=\frac{2 G M}{r}-\frac{\alpha}{r^2}\left(\frac{GM}{r}-\frac{k\tilde r_0}{2}\right)^2$ (see also \cite{giesel2022spherical}). Here the spatial curvature of the dust plays a role in the quantum correction. The effect of this correction is left for our future  study.

The insight of the current work can be manifested by comparing with other works on qOS models. In \cite{bambi2013non}, the collapse of dust and radiation with quantum cosmological corrections was studied. Based on the unconventional properties of the effective matter, it was argued that an event horizon could not form even though there appear apparent horizons during the collapse.
In \cite{achour2020bouncing},  the exact Schwarzschild metric was assumed outside the LQC collapsing ball. By this assumption it was also argued that no BH could form in this model. However, in the current work, the external metric is a priori arbitrary, spherically symmetric and (pseudo) static. The global structure of the exterior spacetime shows that there do exist event horizons during the collapse. It should be noted that the BH metric \eqref{eq:exteriormetric1} was also obtained as a solution to the effective equations in certain LQG spherically symmetric model \cite{kelly2020effective,husain2022fate}. So our result prefers to this spherically symmetric model and indicates its consistency with the LQC model. The  robustness of our calculation for the modified metric  has been confirmed by other considerations \cite{marto2015improved,parvizi2022rainbow,loops22}.  Our results further indicates that the modified metric \eqref{eq:exteriormetric1} is the only spherically symmetric and (pseudo) static metric that fits the collapsing ball of the flat APS model. Moreover, in \cite{husain2022quantum,husain2022fate}, one accepted discontinuity and introduced shock waves to match bouncing interior and non-bouncing exterior,  unlike our model.

There are a few other open issues left by the current work. First, the qOS and the qSC models are considered separately, while a realistic cosmology model containing BHs should  combine the two models together, so that the bubble in the quantum universe should be composed of BHs formed by the collapsing dust. Second,  an alternative dynamics in LQC results in an asymmetric bounce such that a de Sitter cosmos emerges  \cite{yang2009alternative,assanioussi2018emergent,zhang2021loop}. In the asymmetric bouncing spacetime, the modified Friedemann equation changes  its form after the bounce. Thus how to glue a BH with that model is a challenging issue which deserves future investigating.
Last but not the least, the resolution of the singularity in our result is related to the appearance of an inner horizon which could develop instabilities \cite{carballo2021inner}.  Moreover, there might be other quantum gravity phenomena occurring in the high curvature region once Hawking radiation was taken into account \cite{d2021end}. Those phenomena might affect the global structure of the quantum modified spacetime and thus resolve the would-be instabilities. All these issues deserve further investigating.

\begin{acknowledgments}

This  work is supported by the Polish Narodowe Centrum Nauki, Grant No. 2018/30/Q/ST2/00811, and NSFC with Grants No. 11961131013, No. 12165005, No. 11875006, and No. 12275022.

\end{acknowledgments}

\onecolumngrid

\appendix
\section{The junction conditions}\label{App}
 We need to  glue the  APS dust spacetime $\dd s_{\rm APS}^2$ with the static spherically symmetric spacetime 
 \begin{equation}\label{eq:sexteriormetric}
\dd s_{\rm MS}^2=-(1-F(r))\dd t^2+(1-G(r))^{-1}\dd r^2+r^2\dd\Omega^2,
\end{equation}
along radial geodesics therein. Let 
 \begin{align}\label{dustlinesss}
 \tau\mapsto (t(\tau),r(\tau),\theta,\phi)
 \end{align}
 be a geodesic in the spacetime $\dd s_{\rm APS}^2$ and $\tau$ be the proper time.  The functions $r(\tau)$ and $t(\tau)$ satisfy the following identities   
\begin{equation}\label{eq:geodesic0}
\begin{aligned}
-1=&-(1-F(r(\tau)))\dot{t}(\tau)^2+(1-G(r(\tau)))^{-1}\dot{r}(\tau)^2,\\
E=&(1-F(r(\tau)))\dot{t}(\tau),
\end{aligned}
\end{equation}
where the first equation follows from the conservation of the norm  of the vector tangent  to the geodesic, and the  second one is the existence of a constant of motion $E$ resulting from the existence of the Killing vector $\partial/\partial t$.  Simplifying  \eqref{eq:geodesic0}, one gets
\begin{equation}\label{eq:geodesic1}
\begin{aligned}
\dot{t}(\tau)&=\frac{E}{1-F(r(\tau))},\\
\dot{ r}(\tau)^2&=E^2\frac{1-G(r(\tau))}{1-F(r(\tau))}-1+G(r(\tau)).
\end{aligned}
\end{equation}

Along the geodesics (\ref{dustlinesss}) in the spacetime $\dd s_{MS}^2$ and the geodesics
$\tau\mapsto (\tau,{\tilde r}_0,\theta,\phi)$  
in the dust APS spacetime, respectively,  we glue the spacetimes by the identification
\begin{equation}\label{eq:gluing}
(\tau,{\tilde r}_0,\thc,\phc) \sim (t(\tau),r(\tau),\thc,\phc), \ \ \ \ 
\end{equation}
such that the induced metric and the extrinsic curvature are equal on the gluing surfaces that become a single surface of the dusty part of the spacetime. 
By a straightforward calculation, the junction condition leads to
\begin{equation}\label{eq:GF}
\begin{aligned}
a(\tau)\rc_0=r(\tau),\quad E^2\frac{1-G(r(\tau))}{1-F(r(\tau))}=1.
\end{aligned}
\end{equation}
We notice that the first equation rigidly binds the functions $a(\tau)$ and $r(\tau)$, while the second equation equally rigidly binds $F$ and $G$.  Also, the geodesic integrability conditions \eqref{eq:geodesic1} now take a simpler form
\begin{equation}\label{eq:geodesic3}
\begin{aligned}
&\dot{t}(\tau)=\frac{1}{E^2(1-G(r(\tau)))},\\
\dot{ a}(\tau)^2\rc_0^2=&\dot{ r}(\tau)^2=G(r(\tau)).
\end{aligned}
\end{equation}

Combining the lower line of \eqref{eq:geodesic3} with the deformed Friedman equation (2), and the  first equation of (\ref{eq:GF}),  we finally obtain
\begin{equation}\label{eq:Fr'}
G(r(\tau))=\frac{2GM}{r(\tau)}-\frac{\alpha G^2 M^2}{r(\tau)^4}.
\end{equation}    
That makes the function $G$ determined as
\begin{equation}\label{eq:Fr}
G(r)=\frac{2GM}{r}-\frac{\alpha G^2 M^2}{r^4},
\end{equation}
in the range of the values taken by $r$ along te geodesic. The range is $[r_b,\infty)$, where  the lower bound $r_b$ is taken when $\dot r =0$ in (\ref{eq:geodesic3}), hence
\begin{equation}\label{eq:rb}
r_b = a(\tau_c) \rc_0 = \left(\frac{\alpha GM}{2}\right)^{\frac{1}{3}} .  
\end{equation}
 
Applying the second equation of  \eqref{eq:GF}, we thus  get the  static metric  
\begin{equation}
\begin{aligned}
\dd s^2_{\rm MS}=&-\left(1-\frac{2GM}{r}+\frac{\alpha G^2 M^2}{r^4}\right)E^2\dd t^2\\
&+\left(1-\frac{2GM}{r}+\frac{\alpha G^2 M^2}{r^4}\right)^{-1}\dd r^2+r^2\dd\Omega^2.
\end{aligned}
\end{equation}
Without loss of generality, one can choose $E=1$ which is equivalent  to  do the coordinate transformation   $t\to t/E$. Finally, the formula for the static  metric tensor reads
\begin{equation}\label{eq:sexteriormetric1}
\begin{aligned}
\dd s^2_{\rm MS}=&-\left(1-\frac{2GM}{r}+\frac{\alpha G^2 M^2}{r^4}\right)\dd t^2\\
&+\left(1-\frac{2GM}{r}+\frac{\alpha G^2 M^2}{r^4}\right)^{-1}\dd r^2+r^2\dd\Omega^2.
\end{aligned}
\end{equation}

To write the metric \eqref{eq:sexteriormetric1} in a compact form, let us introduce  two quadratic  functions $W(r;\beta)$ and $Y(r;\beta)$,
\begin{equation*}
\begin{aligned}
W(r;\beta):=r^2+\frac{2\beta \sqrt{\alpha}\,r}{\sqrt{(1+\beta)^3(1-\beta)}}+\frac{2\beta^2\alpha}{(1-\beta)(1+\beta)^2}\\
Y(r;\beta):=r^2-\frac{2\beta \sqrt{\alpha}\,r}{\sqrt{(1+\beta)(1-\beta)^3}}+\frac{2\beta^2\alpha}{(1-\beta)^2(1+\beta)},
\end{aligned}
\end{equation*}
where we have replaced  $GM$ by the parameter   $0<\beta<1$ defined by the following relation: 
\begin{equation}
G^2M^2= \frac{4 \beta ^4}{\left(1-\beta ^2\right)^3} \alpha.
\end{equation}
Taking advantage of the two functions, the  metric \eqref{eq:sexteriormetric1} can be rewritten as 
\begin{equation}\label{eq:sexteriormetric2}
\dd s_{\rm MS}^2=-\frac{W(r;\beta)Y(r;\beta)}{r^4}\dd t^2+\frac{r^4}{W(r;\beta)Y(r;\beta)}\dd r^2+r^2\dd\Omega^2,
\end{equation}
The function $W(r;\beta)$ is positive for  all $r>0$ and $0<\beta<1$. Thus, the global  structure of  the spacetime endowed with the metric tensor (\ref{eq:sexteriormetric2}) depends on the number of roots of $Y(r;\beta)$. It turns out that for $0<\beta<1/2$, $Y(r;\beta)$ has no real root, and for $1/2\leq \beta<1$, $Y(r;\beta)$ has two roots which are equal if $\beta=1/2$.


%

\end{document}